\documentclass[twocolumn,english,amssymb,aps,prb,superscriptaddress,tightenlines,notitlepage]{revtex4-2}
\usepackage{multibib}
\newcites{SM}{References}
\usepackage{lmodern}
\usepackage[T1]{fontenc}
\usepackage{color}
\usepackage{babel}
\usepackage{verbatim}
\usepackage{amsmath}
\usepackage{amssymb}
\usepackage{cancel}
\usepackage{graphicx}
\usepackage{esint}
\usepackage[normalem]{ulem}
\usepackage[unicode=true,pdfusetitle,
 bookmarks=true,bookmarksnumbered=true,bookmarksopen=false,
 breaklinks=false,pdfborder={0 0 1},backref=false,colorlinks=true]
 {hyperref}
\hypersetup{
 linkcolor=magenta,urlcolor=blue,citecolor=blue,pdfstartview={FitH},hyperfootnotes=false}

\makeatletter
\usepackage{babel}
\date{\today}
\usepackage{xcolor}
\allowdisplaybreaks

\renewcommand{\big}{\bBigg@\@ne}
\renewcommand{\Big}{\bBigg@{1.5}}
\renewcommand{\bigg}{\bBigg@\tw@}
\renewcommand{\Bigg}{\bBigg@{2.5}}

\newcommand{\biggg}{\bBigg@\thr@@}
\newcommand{\Biggg}{\bBigg@{3.5}}

\newcommand{\before}[1]{\iffalse {#1}\fi}

\newcommand{\ignore}[1]{}

\makeatother

\begin{document}
\title{$H-T$ Phase diagram of CeRh$_{2}$As$_{2}$: Refinement of the parity-switch
scenario}
\author{Changhee Lee}
\email{changhee.lee@otago.ac.nz}
\affiliation{Department of Physics and MacDiarmid Institute for Advanced Materials
and Nanotechnology, University of Otago, P.O. Box 56, Dunedin 9054,
New Zealand}
\author{P. M. R. Brydon}
\email{philip.brydon@otago.ac.nz}
\affiliation{Department of Physics and MacDiarmid Institute for Advanced Materials
and Nanotechnology, University of Otago, P.O. Box 56, Dunedin 9054,
New Zealand}
\begin{abstract}
The superconductivity of CeRh$_2$As$_2$ has drawn attention due to its first-order transition in magnetic fields. At first glance, the multiple superconducting (SC) phases as well as the first-order transition appear consistent with the parity-switch scenario, which emphasizes the role of strong Rashba spin-orbit coupling enabled by the locally non-centrosymmetric crystal structure. However, experimental phase diagrams exhibit notable deviations from this simple picture: thermodynamic measurements reveal a nearly vertical phase boundary of the high-field SC phase despite the orbital depairing effect, while transport measurements show that the initial slope of the high-field SC phase is steeper than that of the low-field SC phase. Here, we show that these discrepancies can be understood by considering the combined effects of nonsymmorphic band structure and coexisting antiferromagnetic order. We demonstrate that the symmetry-enforced electronic structure around the Dirac node and type-II Van Hove saddle points near the X point in the Brillouin zone boundary becomes more anisotropic with increased interlayer hopping amplitude, and this anisotropic band structure naturally accounts for the anomalous initial slope of the odd-parity SC phase. Meanwhile, a phenomenological theory incorporating coexisting antiferromagnetism explains the nearly vertical thermodynamic phase boundary as a consequence of field-enhanced odd-parity superconductivity enabled by the over-suppression of Pauli depairing. 
\end{abstract}
\maketitle

{\it Introduction.}---The magnetic-field-induced first-order transition between two superconducting~(SC) phases is observed in the heavy fermion superconductor CeRh$_{2}$As$_{2}$, and it is widely interpreted by the so-called parity-switch scenario enabled by the crystal's sublattice structure~\citep{Khim2021,Yoshida2012}.
Specifically, the low-field and high-field SC phases correspond to even-parity SC (eSC) and odd-parity SC (oSC) states with the same and opposite signs of the spin-singlet gap on the two sublattices, respectively, in this scenario.
Strong Rashba spin-orbit coupling~(SOC) enabled by the locally non-centrosymmetric structure naturally renders the two SC states nearly degenerate and enables the field-induced first-order transition. However, the onset of a purported antiferromagnetic~(AFM) phase at $T_{0}\approx0.5$~K~and~its~coexistence with both SC phases in CeRh$_{2}$As$_{2}$~\citep{Khim2024uSR,Kibune2022,Ogata2023,Galeski2026} which is not taken into account in the usual parity-switch scenario.

\begin{figure}
\includegraphics[width=0.95\linewidth]{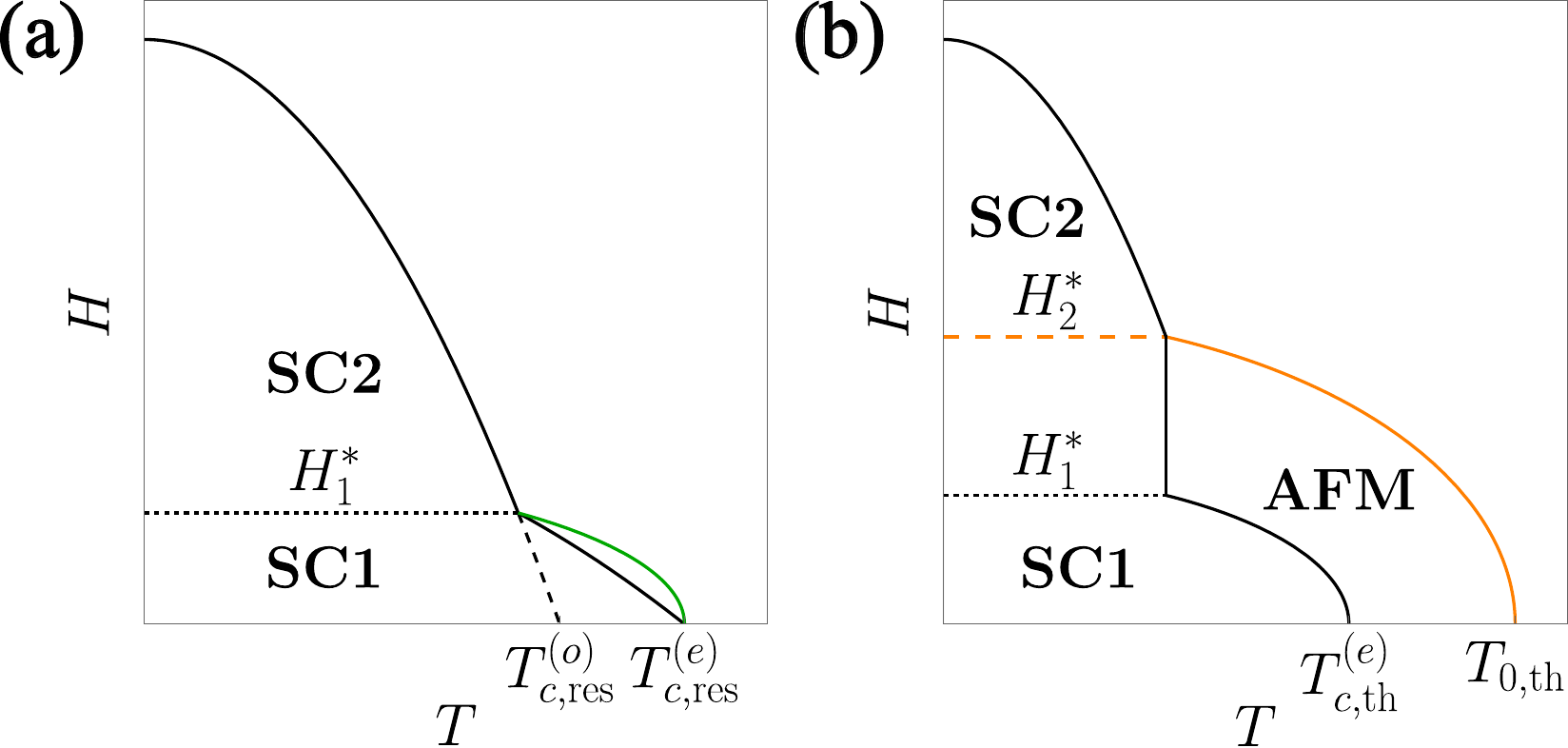}\caption{\label{fig:fig1}Schematic field-temperature ($H-T)$ phase diagrams of CeRh$_{2}$As$_{2}$ in magnetic fields along $c$-axis. (a) Transport phase diagram of the superconducting phases. The dotted line denotes the first order phase transition occurring at $H=H_{1}^{*}$. The dashed line is an extrapolation of the phase boundary in $H>H_{1}^{*}$. The green line depicts a fictitious example of the phase boundary in $H<H_{1}^{*}$ which is consistent with the usual parity-switch scenario. (b) Thermodynamic phase diagram of the superconducting phases along with the AFM phase. $H_{2}^{*}$  indicated by the orange dashed line is the critical field above which the AFM order vanishes.}
\end{figure}

The pronounced differences between magnetic field-temperature ($H$--$T$) phase diagrams obtained from thermodynamic (specific heat, AC susceptibility, etc) and transport (resistivity) measurements~\citep{Semeniuk2023Decoupling,Semeniuk2024Exposing,Pfeiffer2024Pressure,Khanenko2025OutOfPlane}, as schematically illustrated in Fig.~\ref{fig:fig1}, appear to reflect an intricate interplay of the AFM and SC phases. Firstly, the zero-field SC transition temperature appears at a lower temperature in the thermodynamic measurements~($T_{c,{\rm th}}^{({\rm e})}\approx0.3$~K)~than in the transport measurements~($T_{c,{\rm res}}^{({\rm e})}\approx0.4$~K).~However, the first order transition is found to occur almost at the same magnetic field $H^{\ast}_1\approx4$~T in both types of measurements. The onset of the antiferromagnetism is evidenced in thermodynamic measurements, while it is not observed in transport measurements on low quality sample. Recent measurements on higher-quality samples identifies a kink of resistivity as the onset of it~\citep{Semeniuk2023Decoupling}.

Together with the fragility of the antiferromagnetic transition under pressure~\citep{Pfeiffer2024Pressure}, the dependence of the observability of the magnetic transition in transport measurements suggests that the antiferromagnetic phase may be sensitive to disorder-induced strain effects~\citep{Kim2024} as well. Assuming competition between the AFM and SC orders, quenching of the former around disorder provides an explanation for the discrepancy between the transport and thermodynamic phase diagrams: at temperatures where the SC
order is suppressed by the AFM order in the bulk, transport measurements nevertheless detect the SC transition along percolative channels where the AFM is suppressed
due to local disorder, which is supported by the observation that the transport phase diagram remains almost unaltered across the pressures while the AFM order is suppressed~\citep{Pfeiffer2024Pressure}. 


However, even the transport phase diagram is not fully consistent with the usual parity-switch scenario. In particular, the initial slope of the oSC phase, obtained by extrapolating the phase boundary from the kink to zero field, is about 1.5 times steeper than that of the eSC phase. By contrast, conventional Ginzburg-Landau theory formulated within the usual parity-switch scenario predicts that the initial slopes are proportional to the zero-field transition temperatures, reflecting the assumption that the predominantly intrasublattice pairings in the two SC phases experience comparable orbital depairing~\citep{Yoshida2012,Schertenleib2021}. It therefore predicts a steeper initial slope for the low-field eSC phase, as illustrated by the green line in Fig.~\ref{fig:fig1}(a), in contradiction with the transport measurements. More strikingly, the thermodynamic phase diagram exhibits a nearly vertical phase boundary immediately above the first-order transition, despite the expected orbital depairing effect.

In this Letter, we first show that the non-symmorphic symmetries in CeRh$_2$As$_2$ give rise to a highly anisotropic electronic structure around the X point in the first Brillouin zone in the limit of large interlayer hopping amplitude, thereby explaining the anomalously steeper initial slope of the odd-parity SC state.
We then develop a phenomenological Landau theory incorporating SC order and AFM order parameters to explain the nearly vertical phase boundary observed in thermodynamic measurements~\citep{Semeniuk2023Decoupling,Khanenko2025OutOfPlane}. This theory highlight the nearly vertical phase boundary as a consequence of an overcompensation of the negligible Pauli depairing in the odd-parity SC state by the coexisting AFM order.

{\it Transport phase diagram.}---\label{sec:2}The phase transition of a clean type-II superconductor in weak magnetic fields along the $c$-axis can be investigated using a Ginzburg-Landau free energy
$F=\int{\rm d}^{d}\boldsymbol{r}\;{\cal F}$ given by
\begin{align}
{\cal F}= & a(T)|\Delta(\boldsymbol{r})|^{2}+b|\Delta(\boldsymbol{r})|^{4}+K|\boldsymbol{\Pi}\Delta(\boldsymbol{r})|^{2}\label{eq:SCGLmodel}
\end{align}
up to the quartic order of the superconducting order parameter $\Delta(\boldsymbol{x})$.
Here, $\boldsymbol{\Pi}=(\partial_{x},\partial_{y})-\frac{2ie}{\hbar}(A_{x},A_{y})$
are the covariant in-plane derivatives with vector potentials $(A_{x},A_{y})$
and the elementary charge $e$. The coefficient $a(T)=a_{0}(T/T_{c,0}-1)$
of the quadratic term with the transition temperature $T_{c,0}$ at
zero field, and $K=K_{0}T_{c,0}^{-2}$ of the gradient term are associated with the orbital limit of the superconductivity~\citep{Gorkov1964}.
Since we are focusing on the slope of the phase boundary at zero field, the Pauli paramagnetic effect is neglected here as it is less significant than the orbital depairing effect at weak fields.

The standard functional integral technique leads to the following expressions of $a_{0}$ and $K_{0}$ in terms of the parameters of a microscopic fermiology:
\begin{align}
a_{0}&=  \int_{-\Lambda}^{\Lambda}{\rm{d}}uf_{a}(u)\sum_{\lambda}\oint_{\xi_{\boldsymbol{k},\lambda}=2T_{c,0}u}{\mkern-30mu}{\rm{d}}{\cal S}_{\boldsymbol{k}}
\frac{|\Delta_{\boldsymbol{k},\lambda}|^{2}}{|\boldsymbol{v}_{\boldsymbol{k},\lambda}|},\label{eq:a(T)}\\
K_{0}&=2
\,T_{c,0}\int_{-\Lambda}^{\Lambda}{\mkern-5mu}{\rm{d}}uf_{K}(u)\sum_{\lambda}\oint_{\xi_{\boldsymbol{k},\lambda}=2T_{c,0}u}{\mkern-50mu}{\rm{d}}{\cal S}_{\boldsymbol{k}}
|\boldsymbol{v}_{\boldsymbol{k},\lambda}||\Delta_{\boldsymbol{k},\lambda}|^{2},\label{eq:K}
\end{align}
which works in the low temperature limit. The two functions $f_{a}(u)={\rm sech}^{2}(u)$ and $f_{K}(u)=({\rm tanh}\frac{u}{2}-\frac{u}{2}{\rm sech}^{2}\frac{u}{2})/(2u^{3})$ are sharply peaked at the Fermi energy ($u=0$), allowing us to extend the cutoff $\Lambda$ to infinity.
The energy dispersion and the velocity of the $\lambda$ band are denoted by $\xi_{\boldsymbol{k},\lambda}$ and $\boldsymbol{v}_{\boldsymbol{k},\lambda}=\partial_{\boldsymbol{k}}\xi_{\boldsymbol{k},\lambda}$,
respectively. The gap amplitude on the $\lambda$th band is $\Delta_{\boldsymbol{k},\lambda}$;
contributions from interband pairing are neglected as they are negligible
at temperatures much lower than the band separation at the Fermi surface (See the Supplemental Material (SM)~\citep{SM}).

A conventional simplification of Eqs.~(\ref{eq:a(T)}-\ref{eq:K}) is derived by assuming a uniform magnitude of the velocity across the Fermi surface, i.e. $|\boldsymbol{v}_{\boldsymbol{k},\lambda}|=v_{F}$~\citep{Gorkov1964,Cyrot1973}. This leads to the familiar expressions 
\begin{align}
a_{0}= & N_{0}\oint_{{\rm FS}}{\rm{d}}{\cal S}_{\boldsymbol{k}}|\Delta_{\boldsymbol{k},\lambda}|^{2},\label{eq:familara0}\\
K_{0}= & \frac{7\zeta(3)}{8\pi^{2}}N_{0}v_{F}^{2}\oint_{{\rm FS}}{\rm{d}}{\cal S}_{\boldsymbol{k}}|\Delta_{\boldsymbol{k},\lambda}|^{2},\label{eq:familarK0}
\end{align}
where $N_{0}$ is the density of states at the Fermi energy.
Given the standard relation for the upper critical field in a strong
type-II superconductor $H_{{\rm c,2}}(T)=\frac{\hbar}{2e}\frac{a(T)}{K}$~\citep{SzaboPRR2024},
the slope of the critical field at zero magnetic field, which we refer to as the initial
slope, is written as~\citep{Shumei2000,Orlando1979}
\begin{align}
-\partial_{T}H_{{\rm c,2}}|_{T_{c,0}} & =\frac{1}{2\hbar e}\frac{a_{0}T_{c,0}}{K_{0}}\propto\frac{T_{c,0}}{v_{F}^{2}}.\label{eq:Critical_Fields_Slope}
\end{align}
We therefore expect that the initial slope of the upper critical field is
proportional to the zero-temperature critical temperature.

However, the linear dependence on the critical temperature in Eq.~\eqref{eq:Critical_Fields_Slope} appears to fail in CeRh$_{2}$As$_{2}$: the initial slope of oSC, estimated by extrapolating the resistive data just above the first order transition to zero field,
is apparently 1.5 times larger than that of eSC, despite a lower (extrapolated) zero-field critical temperature.
Fitting the data to a semiclassical formula for the upper critical field in a 2D superconductor yields consistent results (See the SM~\citep{SM}).

\begin{figure}
\includegraphics[width=0.95\columnwidth]{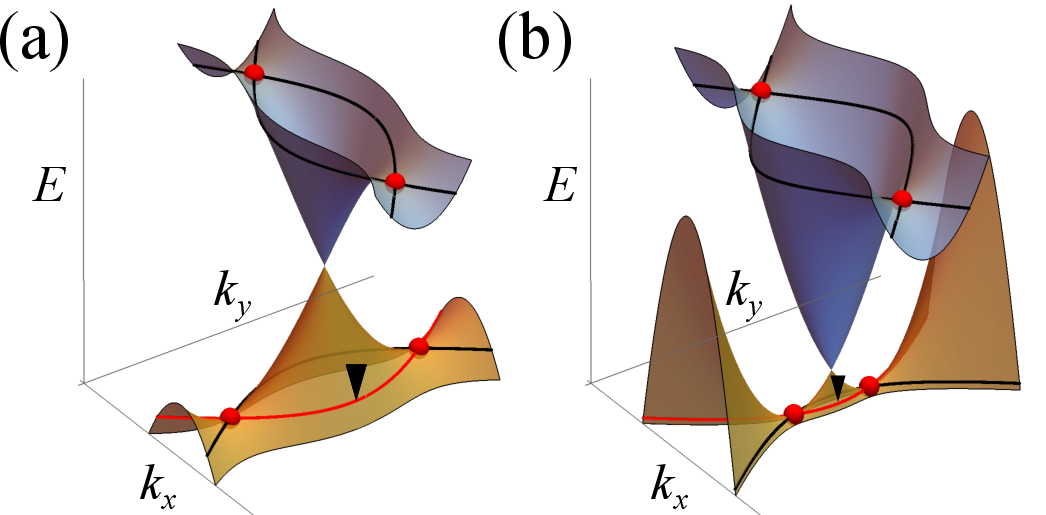}

\caption{\label{fig:fig2}Fermi surfaces of the conduction and the valence
bands. (a) $(t_{\perp},\alpha)=(2,5)t_{0}$ and (b) $(t_{\perp},\alpha)=(5,2)t_{0}$
are used. The red points highlight the Van Hove saddle points. The
curves passing through the red points represent the Fermi surfaces
of the valence (conduction) band at $\mu=\mu_{v}$ ($\mu=\mu_{c}$).
The black arrows mark the midpoints of the red curves and correspond to
the origin of the figures in Fig.~\ref{fig:fig3}. 
}
\end{figure}

It is noteworthy that the linear temperature dependence in Eq.~\eqref{eq:Critical_Fields_Slope} is invalidated when the Fermi velocity has significant anisotropy across the Fermi surface. As we shall show below, such anisotropic band structure can occur if the so-called type-II Van Hove saddle points (VHSs)~\citep{Yao2015,CLee2024}, located away from these time-reversal-invariant momenta, is in the vicinity of the Fermi energy. For CeRh$_2$As$_2$, 
angle-resolved photoemission spectroscopy measurements reported the presence of VHSs around the X($\pi,0$) point of the first Brillouin zone (FBZ)~\citep{XChen2024ARPES,BChen2024ARPES,Wu2024ARPES}.

Nonsymmorphic symmetries of CeRh$_2$As$_2$ indeed enforce VHSs around the X point in the FBZ to be type-II VHSs~\citep{CLee2024} due to a symmetry-enforced Dirac node right at the X point.~The electronic structure around the X point can be conveniently understood using a two-dimensional
$k\cdot p$ Hamiltonian for the X$(\pi,0)$ point of the FBZ~\citep{CLee2024,Suh2023}:
\begin{equation}
H_{\boldsymbol{k}}^{(X)}=\varepsilon_{0,\boldsymbol{k}}I_{4}+t_{\perp}k_{x}\tau_{x}s_{0}+(\alpha_{x}k_{x}s_{y}+\alpha_{y}k_{y}s_{x})\tau_{z}.\label{eq:kpHam}
\end{equation}
Here $\boldsymbol{k}$ is the momentum measured from the X point, while $\tau_\nu$ and $s_\nu$ denote the Pauli
matrices for the sublattice and the spin degrees of freedom, respectively. The nontrivial terms in the Hamiltonian describe interlayer hopping with amplitude $t_{\perp}$, and intralayer spin-orbit coupling (SOC)
with amplitudes $\alpha_{x}$ and $\alpha_{y}$. The interlayer hopping must vanish on the boundary of the FBZ due to the nonsymmorphic symmetries~\citep{Cavanagh2022}, and so it enters into the $k\cdot p$
theory as a term linear in $k_{x}$~\citep{Suh2023}. We choose the trivial part of the Hamiltonian to host a saddle-point dispersion $\varepsilon_{0,\boldsymbol{k}}=-ak_{x}^{2}+bk_{y}^{2}-\mu$, which yields a type-I VHS (i.e. saddle point at the time-reversal-invariant momenta) when the sublattice-spin-nontrivial parts vanish. 

Diagonalizing $H_{\boldsymbol{k}}^{(X)}$, we have the
dispersions  
\begin{equation}
\varepsilon_{\boldsymbol{k},{\rm c(v)}}=\varepsilon_{0,\boldsymbol{k}}\pm\sqrt{(t_{\perp}^{2}+\alpha_{x}^{2})k_{x}^{2}+\alpha_{y}^{2}k_{y}^{2}}.\label{eq:kp_energy}
\end{equation}
where the $+$ and $-$ signs correspond to the conduction band 
and the valence band, respectively;  representative plots of the dispersion are shown in Fig.~\ref{fig:fig2}. The nontrivial terms in Eq.~\eqref{eq:kpHam} have two consequences: the type-I VHS of $\epsilon_{0,\boldsymbol{k}}$ is shifted away from the X point towards the $\Gamma$ ($M$) point in the conduction band (valence band), while the linear momentum-dependence of the nontrivial terms produces a Dirac node at the X point. 
The dispersion around this Dirac node is generally anisotropic. In particular, if the interlayer hopping is larger than the Rashba SOC strength (see Fig.~\ref{fig:fig2}(b)), the band is more dispersive along the $k_x$-direction compared to the $k_y$-direction. This anisotropy is a factor that drives the deviation of the initial slope from the linear temperature dependence in Eq.~\eqref{eq:Critical_Fields_Slope}.

\begin{figure}
\includegraphics[width=1\columnwidth]{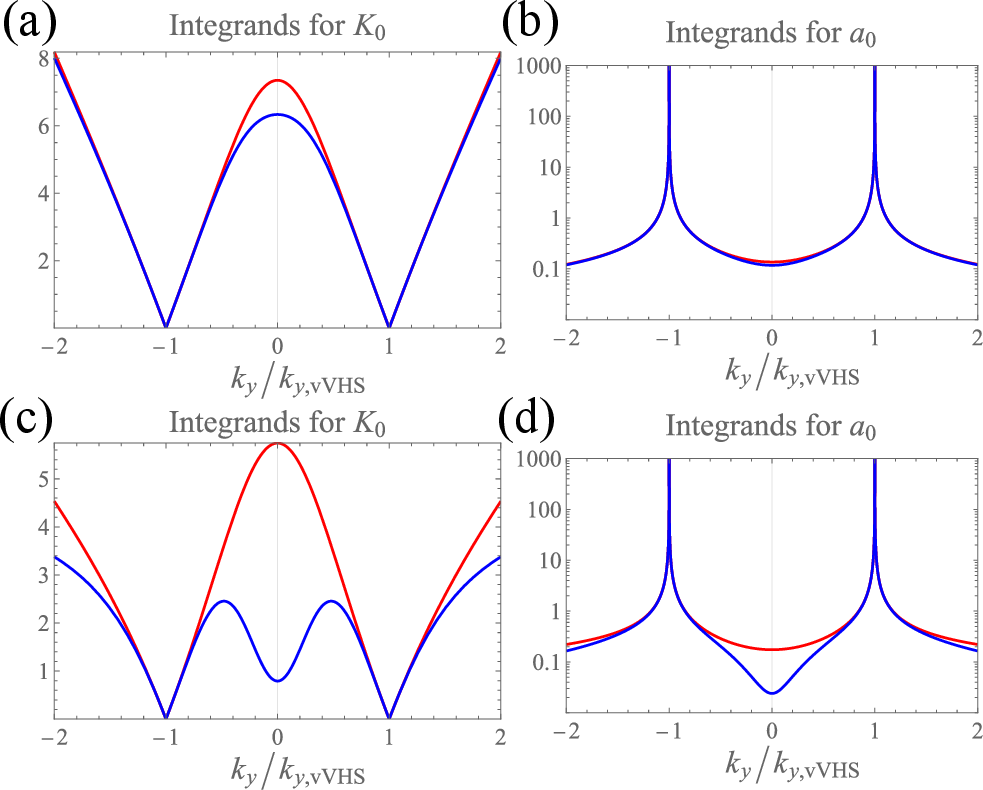}

\caption{\label{fig:fig3}The integrands of $K_{0}^{(p)}$ and $a_{0}^{(p)}$
over the Fermi surface (red curve in Fig.~\ref{fig:fig2}) for the
$B_{{\rm 1g}}$- and $B_{{\rm 2u}}$-type gap functions. (a) and (b)
show the integrands when $(t_{\perp},\alpha)=(2,5)$ and $\mu=\mu_{v}$
in Eq. \eqref{eq:kpHam} with $a=b=1$. The red (blue) line corresponds
to the case with $B_{{\rm 1g}}$($B_{{\rm 2u}}$)-type gap function.
(c) and (d) show the integrands when $(t_{\perp},\alpha)=(5,2)$ in
Eq. \eqref{eq:kpHam} with $a=b=1$. The Fermi surface is parameterized
by using $k_{y}$. $k_{y,{\rm vVHS}}$ denotes the position of the
valence band Van Hove singularity. %
}
\end{figure}



We now introduce superconductivity into the $k\cdot p$ theory in Eq.~\eqref{eq:kpHam}. Theoretical analysis using various methods suggest competing $B_{1g}$ and $B_{2u}$ instabilities in models of CeRh$_2$As$_2$~\citep{Nogaki2022,CLee2024}, which is consistent with a parity-switch scenario between $d_{x^2-y^2}$-wave uniform (e) and staggered (o) singlet pairing states. These are described by the structural form factors $\psi_{\boldsymbol{k}}\tau_{0}s_{0}$
and $\psi_{\boldsymbol{k}}\tau_{z}s_{0}$,
respectively~\citep{Nogaki2022,CLee2024}, where $\psi_{\boldsymbol{k}}$ can be obtained, for example, by expanding $\cos k_{x}-\cos k_{y}$ around the X point. By projecting these form factors onto the $\lambda$th band, we obtain the magnitudes $|\Delta^{({\rm e})}_{\boldsymbol{k},\lambda}|$ and $|\Delta^{({\rm o})}_{\boldsymbol{k},\lambda}|$ of both gap functions around the X point which are given by
\begin{equation}
|\Delta_{\boldsymbol{k},\lambda}^{({\rm e})}|^{2}=2|\psi_{\boldsymbol{k}}|^{2},\;|\Delta_{\boldsymbol{k},\lambda}^{({\rm o})}|^{2}=\frac{2|\psi_{\boldsymbol{k}}|^{2}(\alpha_x^{2}k_x^{2}+\alpha_y^2k_y^2)}{(t_{\perp}^{2}+\alpha_x^2)k_{x}^{2}+\alpha^{2}_yk^{2}_y}.
\end{equation}
Note that $|\Delta^{({\rm o})}_{\boldsymbol{k},\lambda}|$ is the same with $|\Delta^{({\rm e})}_{\boldsymbol{k},\lambda}|$ along the $k_y$ axis, while the interlayer coupling causes $|\Delta^{({\rm o})}_{\boldsymbol{k},\lambda}|$ to be smaller along the $k_y=0$ line than the $k_x=0$ line. Therefore, the transition temperatures of both superconducting states are expected to be more similar when the Fermi level is located at the VHS level of the valence band than the case with the Fermi level at the conduction band VHS level.

Armed with the energy dispersions, the magnitudes of the gap functions, and the vicinity of Van Hove singularity to the Fermi level~\citep{XChen2024ARPES,BChen2024ARPES,Wu2024ARPES}, we evaluate the coefficients $a_{0}^{(p)}$
and $K_{0}^{(p)}$ ($p={\rm e},{\rm o}$) with the Fermi level at the VHS level of the valence band which can keep the transition temperatures of both superconducting states comparable.
Figure~\ref{fig:fig3} shows the the integrands in Eqs.~(\ref{eq:a(T)}-\ref{eq:K}) evaluated over the Fermi surface shown in Fig.~\ref{fig:fig2}. Here we use $\psi_{\boldsymbol{k}}=1$ for simplicity. The integrands of $a_{0}^{({\rm e})}$
and $a_{0}^{({\rm o})}$ presented in Figs.~\ref{fig:fig3}(b)~and~\ref{fig:fig3}(d)
shows that they are almost the same near the VHSs regardless of $t_{\perp}/\alpha$ which is consistent with the identical values of the projected gaps on the zone edges. Considering that VHSs contribute the largest portion to $a_{0}^{(p)}$, this makes us expect that the $a_{0}^{({\rm e})}$ and $a_{0}^{({\rm o})}$ are quantitatively similar even though the contribution from the electronic states far from the VHSs is taken into account.

In contrast, as $|v_{v,\boldsymbol{k}}|=0$ at the VHSs, the primary contribution to the integral in $K_{0}^{(p)}$ from the electronic states of the valence band arises
away from the VHSs. When $t_{\perp}/\alpha\ll1$, the integrands of $K_{0}^{({\rm e})}$ and $K_{0}^{({\rm o})}$ are almost the same as shown in Fig.~\ref{fig:fig3}(a). 
When $t_{\perp}/\alpha\gg1$, however, the velocity becomes anisotropic around the Dirac node and is maximized on the $k_{y}=0$ line where  $|\Delta_{\boldsymbol{k},\lambda}^{({\rm o})}|^{2}$
is minimized. As a result, $K_{0}^{({\rm o})}$ becomes much smaller than $K_{0}^{({\rm e})}$ as shown in Fig.~\ref{fig:fig3}(c). 

\begin{figure}
\includegraphics[width=0.8\columnwidth]{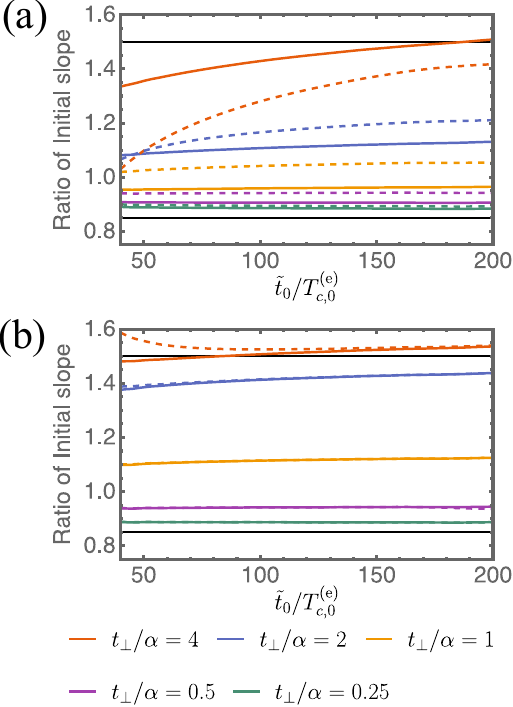}

\caption{\label{fig:fig4}The ratio $(\partial H_{{\rm c,2}}^{({\rm o})}/\partial T_{c,0}^{({\rm o})})/(\partial H_{{\rm c,2}}^{({\rm e})}/\partial T_{c,0}^{({\rm e})})$
of the initial slopes for several $(t_{\perp},\alpha)$ with $t'=0.1t$. (a) shows
the results obtained from the tight-binding model with $\sqrt{t_{\perp}^{2}+\alpha^{2}}=t_{0}$.
(b) shows the results with $\sqrt{t_{\perp}^{2}+\alpha^{2}}=8t_{0}$. The solid lines represent the contribution only from the valence band in Eqs.~(\ref{eq:a(T)}-\ref{eq:K}), while the dashed lines represent the contribution from all bands. Two black lines mark $0.85$ and $1.5$.}
\end{figure}

Consequently, the large ratio $t_{\perp}/\alpha$, making the Dirac cone anisotropic, aggravates the difference between $K_{0}^{(\rm e)}$ and $K_{0}^{(\rm o)}$ while $a_{0}^{({\rm e})}$ and $a_{0}^{({\rm o})}$ are kept relatively similar. This significantly different response of $K_{0}$ and $a_{0}$ against increasing $t_{\perp}/\alpha$ provides an explanation of the large ratio of the initial slopes of the two nearly-degenerate SC phases, which we shall demonstrate using a two-sublattice tight-binding model Hamiltonian
\begin{equation}
H_{\boldsymbol{k}}=\varepsilon_{00}(\boldsymbol{k})I_4+2t_{\perp}\cos\frac{k_{x}}{2}\cos\frac{k_{y}}{2}\tau_{x}s_{0}+\tau_z(\vec{\alpha}_{\boldsymbol{k}}\cdot\vec{s}),\label{eq:TBHam}
\end{equation}
with $\varepsilon_{00}(\boldsymbol{k})=-2t_{0}(\cos k_{x}+\cos k_{y})-4t'(\cos k_{x}\cos k_{y}+1)-\mu$, and $\vec{\alpha}_{\boldsymbol{k}}=\alpha(\sin k_y,-\sin k_x,0).$
When expanded around the X point of the FBZ, $H_{\boldsymbol{k}}$
is turned into $H_{\boldsymbol{k}}^{(X)}$ in Eq.~\eqref{eq:kpHam}
with understanding $a+b=2t_{0}$ and $b-a=4t'$. The hopping parameters are scaled to keep the bandwidth $W=8\tilde{t}_0$ fixed to $8$ so that $\tilde{t}_0$ reduces to $t_0$ in the absence of $t_\perp$ and $\alpha$ for $|t'|<|t_0|/2$. This tight-binding model let us compare the $s$-wave and the $d$-wave gap functions which are not distinguished in the continuum model.

Using $a_0^{(p)}$ and $K_0^{(p)}$ obtained from the tight-binding Hamiltonian for different $(t_{\perp},\alpha)$, we obtain the ratio of the initial slopes of two SC states as a function of the inverse superconducting transition temperature $T_{{\rm c,0}}^{({\rm e})}$ of the eSC phase, which is presented Figure~\ref{fig:fig4}. 
As for the \textit{bare} transition temperature of oSC from the staggered gap function, $T_{{\rm c,0}}^{({\rm o})}=0.85T_{{\rm c,0}}^{({\rm e})}$ is used. The solid lines depict the result obtained by taking the contribution from the valence band in Eqs.~(\ref{eq:a(T)}-\ref{eq:K}) while ignoring the presence of the other band. The results obtained by taking the effect of the multiple bands into account are drawn by the dashed lines (See the SM~\citep{SM}). 
The contribution from the conduction band becomes negligible if temperature is much lower then the separation between the Fermi level and the conduction band minimum. Thus, the solid and the dashed lines approach to each other with the decreasing temperature, implying the dominance of the contribution from the VHS in the valence band to, especially, $a_0^{(p)}$. 

As shown in Fig.~\ref{fig:fig4}, the ratio of the slopes saturates at a value around $T_{{\rm c,0}}^{({\rm o})}/T_{{\rm c,0}}^{({\rm e})}\sim0.85$ when $t_{\perp}/\alpha\ll1$ in accordance with Eq.~\eqref{eq:Critical_Fields_Slope}. Therefore, a superconducting phase diagram is expected where the initial slope of oSC lower than that of eSC, similar to the results presented in Ref.~\citep{Schertenleib2021} for $K^{\rm (e)}_{0}=K^{\rm (o)}_{0}$. Contrastingly, the ratio of the initial slopes reaches to a larger value than $T_{{\rm c,0}}^{({\rm o})}/T_{{\rm c,0}}^{({\rm e})}$ with increasing $t_{\perp}/\alpha$ which is consistent with the transport phase diagram. This demonstrates that the anisotropy of the dispersion around the Dirac node and the magnitude of the staggered gap function can yield the ratio of the initial slope close to the value estimated from the transport phase diagram.

Our analysis also reveals that the role of the type-II VHSs is more prominent in the $d$-wave superconducting state than in the $s$-wave superconducting state. This is because the nodes of the $d$-wave gap on the Brillouin zone diagonals suppress the contribution from the region around the nodes of the gap function and make the contribution from the region around the VHSs near the X point more significant to $a_0$ and $K_0$. Meanwhile, the entire Fermi surface contributes a rather uniform contribution to $a_0$ and $K_0$ in the case of an $s$-wave gap function, and it reduces the significance of the contribution from the electronic states near the X point. As a result, the ratio of the initial slopes from the $s$-wave gap functions saturates at a value lower than the case with $d$-wave gap functions in general as shown in the SM~\citep{SM}. 

{\it Coexisting AFM order.}---\label{sec:3}The nearly vertical superconducting phase boundary in the regime $H^{}_1 < H_z < H^{}_2$ (Fig.~\ref{fig:fig1}(b)) suggests that superconductivity in this region is strongly modified by the coexisting antiferromagnetic order, consistent with the recovery of conventional behavior for $H > H^{*}_2$. To describe the interplay between the two order parameters, we consider a phenomenological Landau free energy within the lowest Landau level approximation:
\begin{align}
F^{(p)}= & F^{(p)}_{\Delta}+F_{{\rm AFM}}+g^{(p)}|\Delta^{(p)}|^{2}S^{2},\label{eq:CoexistingLandau}\\
F^{(p)}_{\Delta}= & a^{(p)}(T)|\Delta^{(p)}|^{2}+b^{(p)}|\Delta^{(p)}|^{4}\\
 & +(\chi_{{\rm O}}^{(p)}H_{z}+\chi_{{\rm P}}^{(p)}H_{z}^{2})|\Delta^{(p)}|^{2},\nonumber \\
F_{{\rm AFM}}= & a_{{\rm A}}S^{2}+b_{{\rm A}}S^{4}+\chi_{{\rm A}}H_{z}^{2}S^{2},
\end{align}
where $\Delta^{(p)}$ ($p=\rm e,o$) denote an even-parity and an odd-parity SC order parameters and $S$ denotes the AFM order parameter. The coefficients of the quadratic terms are given by $a^{(p)}(T)=T/T_{c,0}^{(p)}-1$ and $a_{{\rm A}}(T)=T/T_{0}-1$. $T_{c,0}^{(p)}$ is the bare SC transition temperature at zero field in the absence of the AFM order. $\chi_{{\rm P}}^{(p)}$ is a coefficient associated with the Pauli depairing effect. The orbital depairing effect is simply captured through the term linear in $H_{z}$. Here we allow $\chi_{{\rm O}}^{(e)} \neq \chi_{{\rm O}}^{(o)}$ in line with our microscopic analysis. $\chi_{{\rm A}}$ describes the suppression
of the AFM order in magnetic fields. $g^{(p)}$ is the coupling between the SC and the AFM order parameters.

The intrinsic properties of the SC states can be estimated by making use of the transport measurement. Ignoring the AFM order parameter, we obtain two ground states from $F^{(p)}_{\Delta}$ whose boundaries against the normal phase are given by 
\begin{equation}
T^{(p)}_{c}=T^{(p)}_{c,0}(1-\chi^{(p)}_{\rm O}H_z-\chi^{(p)}_{\rm P}H_z^2).\label{eq:PureTc}
\end{equation}
$T^{(p)}_{c,0}$, $\chi^{(p)}_{\rm P}$, and $\chi^{(p)}_{\rm O}$, which are intrinsic to SC order parameters, are obtained by using the phase boundary determined by resistivity in $H_z<10$~T as the Landau free energy in Eq.~\eqref{eq:CoexistingLandau} is not valid for too strong fields. In particular, $\chi^{(\rm o)}_{\rm P}$ is found to be very small in line with the identification of the high-field SC state as an odd-parity SC state in the parity-switch scenario. The resultant phase boundaries $T^{(p)}_{c}(H_z)$ are depicted by the black dashed lines, while the black dashed line around $H_z=4$~T, representing the first order transition between two SC states, is determined by comparing $F^{({\rm e})}_{\Delta}$ and $F^{({\rm o})}_{\Delta}$. 

\begin{figure}
\includegraphics[width=0.85\columnwidth]{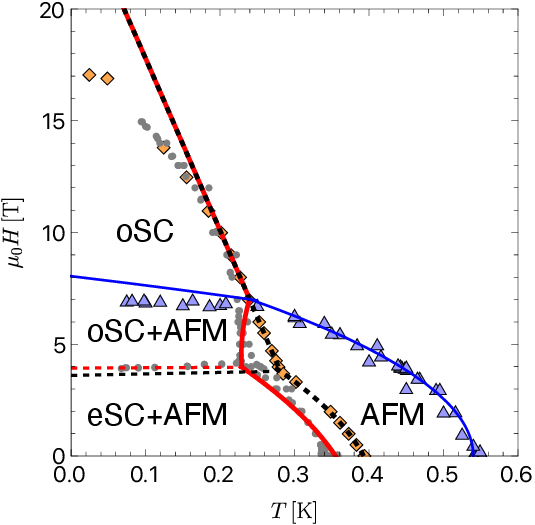}
\caption{\label{fig:fig5}Phase diagram obtained from Eqs.~(\ref{eq:PureTc}-\ref{eq:Tc}). The orange diamonds represent the resistivity measurement in Ref.~\citep{Semeniuk2024Exposing}, while the other symbols correspond to the thermodynamic data presented in Ref.~\citep{Khanenko2025OutOfPlane}. The black dashed lines represent the phase boundaries between the normal phase and the eSC and oSC states, and the first order transition line between two SC states in the absence of the AFM order. The red and blue lines are the phase boundaries obtained by using Eqs.~(\ref{eq:TAFM}-\ref{eq:Tc}) which take account of the AFM order. The red dashed line is the first order transition line between two SC states in the presence of the AFM order.~Here we use~$(T^{(\rm e)}_{c,0},\chi^{(\rm e)}_{\rm O},\chi^{(\rm e)}_{\rm P})=(0.393,0.04,0.0093)$,~$(g^{(\rm e)},b^{(\rm e)})=b_{\rm A}(0.561,0.710)$,~$(T^{(\rm o)}_{c,0},\chi^{(\rm o)}_{\rm O},\chi^{(\rm o)}_{\rm P})=(0.330,0.0385,3.3\times10^{-4})$, $(g^{(\rm o)},b^{(\rm o)})=b_{\rm A}(0.755,0.971)$,~and~$(T_{0},\chi_{\rm A})=(0.54,0.011)$.
}
\end{figure}

With $T_{c,0}^{(\rm{o})}<T_{c,0}^{(\rm{e})}<T_{0}$, the Landau free energies $F^{(p)}$ in Eq.~\eqref{eq:CoexistingLandau} can yield four ground states relevant to CeRh$_2$As$_2$: (1) an AFM phase where only the AFM order parameter is finite (AFM phase), (2) an even-parity SC phase coexisting with AFM ordering (eSC+AFM phase), (3) an odd-parity SC phase coexisting with AFM ordering (oSC+AFM phase), and (4) a pure odd-parity SC phase (oSC phase). Each phase is marked in Figure~\ref{fig:fig5}.

The phase boundary $T_A(H_z)$ of the oSC+AFM phase is given by
\begin{equation}
\frac{T_{{\rm A}}}{T_{0}}=\frac{1-\frac{g^{(\rm o)}}{2b^{(\rm o)}}(1-\chi_{{\rm O}}^{(\rm o)}H_{z})-(\chi_{{\rm A}}-\frac{g^{(\rm o)}\chi_{{\rm P}}^{(\rm o)}}{2b^{(\rm o)}})H_{z}^{2}}{1-\frac{g^{(\rm o)}}{2b^{(\rm o)}}\frac{T_{0}}{T_{c,0}^{(\rm o)}}},\label{eq:TAFM}
\end{equation}
while the boundary of the AFM phase is obtained by taking $g^{({\rm o})}=0$. These two boundaries are depicted by blue lines in Fig.~\ref{fig:fig5}. The boundary of the phase where the the AFM order and the SC order $\Delta^{(p)}$ are finite is given by 
\begin{equation}
T_{c}^{(p)}(H_{z})=T_{c}^{(p)}(0)(1-\tilde{\chi}_{{\rm O}}^{(p)}H_{z}-\tilde{\chi}_{{\rm P}}^{(p)}H_{z}^{2}),\label{eq:Tc}
\end{equation}
where $T_{c}^{(p)}(0)\equiv T^{(p)}_{c,0}(1-\frac{g^{(p)}}{2b_{\rm A}})/(1-\frac{g^{(p)}}{2 b_{\rm A}}\frac{T_{c,0}}{T_{0}})$ is the superconducting transition temperature at zero field in the presence of the AFM order. $\tilde{\chi}^{(p)}_{\rm O}$ and $\tilde{\chi}^{(p)}_{\rm P}$ capture the effective orbital and Pauli depairing in the presence of the coexisting AFM order, respectively, and are given by
\begin{equation}
\tilde{\chi}^{(p)}_{\rm O}=\frac{1}{r}\chi^{(p)}_{\rm O},\quad
\tilde{\chi}^{(p)}_{\rm P}= \frac{1}{r}\big({\chi^{(p)}_{\rm P}-\frac{g^{(p)}}{2b_{\rm A}}\chi_{\rm A}}\big),\label{eq:effectivePauli}
\end{equation}
with $r=1-g^{(p)}/(2 b_{\rm A})$. The curved red lines Fig.~\ref{fig:fig5} represent $T_c^{(p)}(H_z)$ of coexisting phases fitted to the thermodynamic data. The straight red line corresponds to the boundary of the oSC phase obtained by taking $g^{({\rm o})}=0$ in Eq.~\eqref{eq:Tc}

The parameters $T_0$, $g^{(p)}/b_{\rm A}$, and $b^{(p)}/b_{\rm A}$ are extracted from the thermodynamic phase boundary data in Ref.~\citep{Semeniuk2024Exposing}, depicted by gray circles and the blue triangles in Fig.~\ref{fig:fig5}, in the way that the critical point where $T^{(\rm o)}_{c}(H_z)$ in Eq.~\eqref{eq:Tc} and $T_{A}(H_z)$ in Eq.~\eqref{eq:TAFM} intersects on the line by $T^{(\rm o)}_{c}(H_z)$ in Eq.~\eqref{eq:PureTc}. The condition for the coexistence of SC and AFM orders $2 b_{\rm A} b^{({p})}-(g^{(p)})^2>0$ is satisfied by the extracted parameters, and the resultant phase boundaries agree overall well with the experimental data for $H_z<10$~T. Though the boundary (blue line) at $H_z \approx 7$~T  between the oSC+AFM phase and the oSC phase appears to deviate slightly from the experimental data, its negative slope is still consistent with the thermodynamic constraint~\citep{Yip1991}.

It is noteworthy that the boundary between the oSC+AFM phase and the AFM phase, depicted by a red line in $4<H_z<7$~T in Fig.~\ref{fig:fig5}, exhibits a field-enhanced behavior. This can be understood as an overcompensated Pauli depairing effect from Eq.~\eqref{eq:effectivePauli}. It occurs because the staggered SC pairing is so robust against the Pauli depairing to ensure that the coefficient $\tilde{\chi}^{({\rm o})}_{\rm P}$ for effective Pauli depairing effect is generally negative indicating field-enhancement. Consequently, our phenomenological analysis suggests that the nearly vertical boundary of the AFM+oSC phase in the thermodynamic phase diagram originates from the overcompensation of the Pauli depairing effect by the AFM order.
{\it Conclusion.}---
We have investigated the anomalous thermodynamic and transport $H$--$T$ phase diagrams of CeRh$_2$As$_2$ using complementary microscopic and phenomenological approaches. We showed that the steeper initial slope of the odd-parity SC phase in the transport phase diagram can arise from the anisotropic electronic structure near the X point of the first Brillouin zone in the limit of strong interlayer hopping. Though this limit is opposite to the preconditions of the usual parity-switch scenario, the vicinity of the Fermi level to the energy level of Van Hove saddle points on the Brillouin zone boundary still supports near degeneracy of two superconducting pairing channels.
We further demonstrated that coexisting antiferromagnetic order renormalizes the Pauli depairing effect on superconductivity and can produce field-enhanced behavior of the phase boundary of a odd-parity superconducting state, which is robust against Pauli depairing. This provides a natural explanation for the nearly vertical phase boundary observed in thermodynamic measurements. These results highlight the essential roles of the nonsymmorphic electronic structure and antiferromagnetic order in understanding the superconducting phase diagram of CeRh$_2$As$_2$ beyond the conventional parity-switch scenario.

{\it Acknowledgments.}---
CL and PMRB were supported supported by the Marsden Fund Council from Government funding, managed by Royal Society Te Ap\={a}rangi, Contract No. UOO2222. We acknowledge useful discussions with Daniel F. Agterberg, Youchi Yanase, Kenji Ishida, Shunsaku Kitagawa, and particularly thank to Elena Hassinger, Konstantin Semeniuk and Meike Pfeiffer for sharing of experimental data in Ref.~\citep{Khanenko2025OutOfPlane}.

{\it Data Availability.}---The data that support the findings of this article are openly available~\citep{data}.
\bibliography{ref}

\begin{thebibliography}{28}
\bibitem [{dat()}]{data}%
  \BibitemOpen
  \href@noop {} {\ }\bibinfo {note} {[data url] will be added when this
  manuscript published.}\BibitemShut {Stop}%
\end{thebibliography}

\vfill{}


\end{document}